%% file: samplepaper.tex
\documentclass[runningheads]{llncs}
\usepackage[T1]{fontenc}
\usepackage{graphicx}
\usepackage{mdframed}
\usepackage{amsmath}
\usepackage{thmtools}  
\usepackage{thm-restate}
\usepackage{todonotes}

\usepackage{hyperref}
\usepackage{cleveref}

\usepackage[ruled,vlined]{algorithm2e}
\SetKwInput{KwInput}{Input}                
\SetKwInput{KwOutput}{Output} 

\crefname{algocf}{Algorithm}{Algorithms}
\Crefname{algocf}{Algorithm}{Algorithms}

\SetAlCapSkip{0.5em} 

\input{commands}
\begin{document}
\title{Exact Matching and Top-k Perfect Matching Parameterized by Neighborhood Diversity or Bandwidth}
\titlerunning{EM and TkPM Parameterized by Neighborhood Diversity or Bandwidth}
%
\author{Nicolas El Maalouly \and Kostas Lakis}
\authorrunning{N. El Maalouly \and K. Lakis}
%
\institute{ETH Zurich, Zurich, Switzerland\\ \email{\{nicolas.elmaalouly,konstantinos.lakis\}@inf.ethz.ch}}
\maketitle              
\begin{abstract}

The Exact Matching (EM) problem asks whether there exists a perfect matching which uses a prescribed number of red edges in a red/blue edge-colored graph. While there exists a randomized polynomial-time algorithm for the problem, only some special cases admit a deterministic one so far, making it a natural candidate for testing the P=RP hypothesis. A polynomial-time equivalent problem, Top-k Perfect Matching (TkPM), asks for a perfect matching maximizing the weight of the $k$ heaviest edges.

We study the above problems, mainly the latter, in the scenario where the input is a blown-up graph, meaning a graph which had its vertices replaced by cliques or independent sets. We describe an FPT algorithm for TkPM parameterized by $k$ and the neighborhood diversity of the input graph, which is essentially the size of the graph before the blow-up; this graph is also called the prototype. We extend this algorithm into an approximation scheme with a much softer dependency on the aforementioned parameters, time-complexity wise. Moreover, for prototypes with bounded bandwidth but unbounded size, we develop a recursive algorithm that runs in subexponential time. Utilizing another algorithm for EM on bounded neighborhood diversity graphs, we adapt this recursive subexponential algorithm to EM.

Our approach is similar to the use of dynamic programming on e.g.\ bounded treewidth instances for various problems. The main point is that the existence of many disjoint separators is utilized to avoid including in the separator any of a set of ``bad'' vertices during the split phase. 

\end{abstract}
\section{Introduction}
\input{introduction}

\section{Preliminaries}
\input{preliminaries}

\section{Top-k Perfect Matching in Graphs of Bounded Neighbourhood Diversity}\label{sec:bounded_neighborhood_diversity}
\input{bounded_neighborhood_diversity}

\section{Going Further: A Recursive Approach}\label{sec:recursive}
\input{recursive_algorithm}

\section{Adapting the Recursive Algorithm to Exact Matching}\label{sec:recursive_EM}
\input{recursive_EM}

\begin{credits}
\subsubsection{\ackname} The authors wish to thank Andor Vári-Kakas for his insightful help during the initial stages of the project. 

\end{credits}
%
%
%
\bibliographystyle{splncs04}
\bibliography{bibliography}

\appendix
\clearpage
\input{appendix.tex}





\end{document}

%% file: commands.tex
\newcommand{\computationalproblem}[3]{%
  \begin{mdframed}[linewidth=1pt]
    \textbf{#1} \\
    \textbf{Input:} #2 \\
    \textbf{Task:} #3
  \end{mdframed}
}

\newcommand{\ebm}[1]{}

\newcommand{\weightededge}[3]{\ensuremath{\left(\{#1, #2\}, #3\right)}}
\newcommand{\topkvalue}[1]{\ensuremath{|#1|_k}}
\newcommand{\prototype}{\ensuremath{\mathcal{P}}}
\newcommand{\tightedges}{\ensuremath{\mathcal{T}}}
\newcommand{\bandwidth}{\ensuremath{\phi}}
\newcommand{\bigseparator}{\ensuremath{\mathcal{S}}}
\newcommand{\bigO}{\ensuremath{\mathcal{O}}}
\newcommand{\ASfactor}{\ensuremath{\alpha}}

\renewcommand{\todo}[1]{}

\renewcommand{\epsilon}{\varepsilon}

%% file: introduction.tex
The concept of Perfect Matchings (PM) is central to the study of a long catalog of graph-theoretic algorithmic problems. By deciding the existence of a PM or finding one which maximizes (or minimizes) the total edge weight, one can design efficient algorithms for such problems, since the former tasks are solvable in polynomial time \cite{paths_trees_flowers}.

An interesting twist one can impose on the PM problem is the following. Given an edge-coloring of the input graph $G$ using two colors (say red and blue), one can ask whether there exists a PM of $G$ which includes precisely $k$ red edges, where $k$ is also part of the input. We call this problem \textbf{Exact Matching (EM)}.

\computationalproblem{Exact Matching (EM)}{A graph $G$ with edges colored red/blue and an integer $k$.}{Decide if there exists a PM $M$ of $G$ such that $M$ has exactly $k$ red edges.}

This seemingly benign modification to the PM problem was conceived by Papadimitriou and Yannakakis \cite{Papadimitriou_Yannakakis_EM} whilst trying to investigate the effect of restricting the shape of the output tree in the minimum weight spanning tree problem. Specifically, this restriction can be expressed as a graph class which the output tree must fall into. For example, if this class is the one containing all paths, then the problem is at least as hard as deciding the existence of a Hamiltonian path, which is \textbf{NP}-hard. Interestingly, Papadimitriou and Yannakakis showed that for a specific class (called \textit{double 2-stars}) the resulting constrained spanning tree problem is polynomial-time equivalent to EM.

At the time of its conception, EM was conjectured to be \textbf{NP}-hard. A few years later, though, Mulmuley, Vazirani and Vazirani \cite{EM_poly_random} used a polynomial identity testing approach based on the Schwartz-Zippel Lemma \cite{Scharwtz,Zippel} along with the \textit{isolation lemma} they developed to solve the problem in randomized polynomial time. Therefore, it was shown that it is quite unlikely for EM to be \textbf{NP}-hard after all. Even after so many years however, no deterministic algorithm is known for the general case of EM. This borderline placement of the problem in terms of computational difficulty is what makes it interesting. Specifically, it pertains to the long standing question of whether \textbf{P=RP}, being one of the very few natural problems suitable to test this question.

In an effort to study EM from a different angle, El Maalouly introduced the \textbf{Top-k Perfect Matching (TkPM)} problem \cite{algos_and_related_problems}, which is the main focus of this work. In TkPM, an edge-weighted (as opposed to edge-colored) graph $G$ is given as input along with a positive integer $k$. The task is to compute a matching $M$ which maximizes the total weight of the $k$ heaviest edges.

\computationalproblem{Top-k Perfect Matching (TkPM)}{A graph $G$, a weight function $w$ over the edges and an integer $k$.}{Compute a PM $M$ of $G$ which maximizes $|M|_k$ among all PMs of $G$, where $|M|_k$ refers to the total weight of $M$'s $k$ heaviest edges.}

For $k = \frac{n}{2}$, the problem reduces to the classical maximum weight perfect matching problem. The investigation in question \cite{algos_and_related_problems,top_k_note} showed that TkPM (with polynomial weights, which we also assume) is in fact polynomial-time equivalent to EM, making TkPM another valid candidate for the \textbf{P=RP} question. The main difference between EM and TkPM is that the latter can be naturally tackled as an approximation task. Conceptually similar optimization objectives had already been studied for problems such as $k$-clustering and load balancing \cite{load_balancing}.
\paragraph*{Previous work on TkPM.}
El Maalouly~\cite{algos_and_related_problems} studied TkPM both from an exact and an approximation viewpoint. All algorithms mentioned below are deterministic. For general graphs, an algorithm achieving a $0.5$ approximation ratio was given. This was specialized for bipartite graphs, yielding an approximation ratio that tends to $0.8$ polynomially with respect to the number of vertices.

For the exact algorithms, the independence number $\alpha$ and bipartite independence number $\beta$ of the graph was utilized. The bipartite independence number is defined in a similar way to the independence number. For a given bipartition $A, B$ of $V(G)$, the bipartite independence number $\beta$ is the largest number such that there exists an independent set in $G$ using $\beta$ vertices from $A$ and another $\beta$ vertices from $B$, yielding a total of $2\beta$ vertices. Given these parameters, exact algorithms were proposed that run in \textit{Fixed Parameter Tractable (FPT)} time with respect to both $k, \alpha$ in the general case and $k, \beta$ in the bipartite case.

\paragraph*{Our contributions.}
In this work we mainly concern ourselves with blown-up graphs. By blowing up a graph we mean the process of repeatedly taking one original vertex and replacing it with a clique or an independent set (which we call a \textit{blob}) of arbitrary size. These newly created vertices inherit the same neighborhood relationships as the original vertex for vertices outside the blob.

\ebm{
First, in Section \ref{sec:exact-b-matching}, we show a reduction of a conceptually ``blown-up'' version of EM to EM itself. We call this problem \textit{Exact b-Matching (EbM)}. It is defined in a similar fashion as EM, but instead of a PM, we are looking for a set of edges which satisfies a pre-described incidence cardinality constraint for each vertex. In the PM scenario, this cardinality constraint asks that each vertex is incident to exactly one edge. We generalize this such that each vertex demands a specific number of edges incident to it. This is inspired by the well-known b-Matching problem, which is simply the uncolored version of EbM. The gadgets we use in our reduction are essentially blow-ups of the original graphs.
\begin{restatable}{theorem}{exactbmatching}
\label{thm:exactbmatching}
EbM $\equiv_{p}$ EM.
\end{restatable}}

In Section \ref{sec:bounded_neighborhood_diversity}, we focus on TkPM for graphs which are blow-ups of constant size graphs. More accurately, we design an exact algorithm that runs in FPT time with respect to $k$ and the \textit{neighborhood diversity} parameter. Intuitively, a graph has neighborhood diversity $\gamma$, if its vertices can be partitioned into $\gamma$ sets such that the neighborhood of a vertex only depends on which set of the partition includes it. We formally define this in~\Cref{def:bounded-neighbourhood-diversity}.

\begin{restatable}{theorem}{basiccorrectness}
\label{the:basic_correctness}
    Algorithm \ref{alg:basic} computes a Perfect Matching maximizing the top-k objective over all Perfect Matchings of the input graph $G$, if one exists. Otherwise it returns the empty set.
\end{restatable}
\begin{restatable}{theorem}{basicruntime}
\label{the:basic_runtime}
    Algorithm \ref{alg:basic} runs in time $\bigO\left(\binom{2k + \gamma - 1}{\gamma - 1}f(n) \right)$, where $f(n)$ is a polynomial function. In other words, Algorithm \ref{alg:basic} runs in FPT time when parameterized by both $\gamma$ and $k$.
\end{restatable}

We modify the previous exact algorithm slightly and derive an approximation scheme whose complexity is governed by the accuracy required, $k$ and $\gamma$. However, the dependency on the latter two is much lighter.

\begin{restatable}{theorem}{approxcorrecntess}
\label{the:approx_correctness}
    Algorithm \ref{alg:approx} computes a Perfect Matching approximately maximizing the top-k objective within a $(1 - \epsilon)$ factor over all Perfect Matchings of the input graph $G$, if one exists. Otherwise it returns the empty set.
\end{restatable}

\begin{restatable}{theorem}{approxruntime} \label{the:approx_runtime}
    Algorithm \ref{alg:approx} runs in time $\bigO\left(\left(\frac{\log_{2}k}{\log_{2}\left(1/(1-\epsilon)\right)}\right)^{\gamma^2} f(n)\right)$, where $f(n)$ is a polynomial function depending only on $n$.
\end{restatable}

Going back to exact algorithms, in Section \ref{sec:recursive} we study how the structure of the initial graph (before the blow-up) can help us solve the problem efficiently. We show that if the initial graph has bounded \textit{bandwidth}, then we can solve TkPM exactly in the blown-up graph in subexponential time. The bandwidth of a graph has been treated as both a quantity to compute and as a parameter to derive efficient algorithms as we do here~\cite{saxe1980dynamic,golovach2011bandwidth,monien1981bounding,kaplan1996pathwidth,beisegel2025graph,zabih1990some}. We do so by exploiting the fact that bounded-bandwidth graphs have many \textit{disjoint}, balanced and small separators. This is important because with our recursive approach there are blobs which we cannot afford to use to split the initial graph for the recursion step, so we must find a separator which contains none of them. We are fortunately always able to find one such ``free'' separator, as long as the graph is large. When it is small, we treat this as a base case, using Algorithm \ref{alg:basic}.

\begin{restatable}{theorem}{recursivecorrectness}\label{the:recursive_correctness}
    Algorithm \ref{alg:recursive} computes a Perfect Matching maximizing the top-k objective over all Perfect Matchings of the input graph $G$, if one exists. Otherwise it returns the empty set. 
\end{restatable}
\begin{restatable}{theorem}{recursiveruntime}\label{the:recursive_runtime}
    Algorithm \ref{alg:recursive} runs in time at most $2^{\bigO\left(\bandwidth^2 \sqrt{n} \log^2{n}\right)}$ when the input prototype $\prototype$ has bandwidth $\bandwidth$.
\end{restatable}

In Section \ref{sec:recursive_EM}, we discuss how one can adapt this algorithm to work for EM. We employ an existing algorithm for EM in bounded neighborhood diversity graphs~\cite{maalouly2024exactmatchingproblemdense} as a base case. Since this algorithm runs in XP time and not FPT as our (base case) algorithm for TkPM, we have to slightly modify our recursive approach details (the definition of which blobs we can/cannot use for splitting) and that results in a worse, but still subexponential runtime.

\begin{restatable}{theorem}{recursiveEM}\label{the:recursiveEM_runtime}
    There exists an algorithm that solves EM and runs in time at most $2^{\bigO\left(\bandwidth^2 n^{\frac{12}{13}} \log^2{n}\right)}$ when the input prototype $\prototype$ has bandwidth $\bandwidth$.
\end{restatable}



%% file: preliminaries.tex
An edge from $u$ to $v$ with weight $w$ is denoted by $\weightededge{u}{v}{w}$. For a matching $M$, we use $|M|$ to denote the sum of the weights of the edges in $M$. Similarly, we use $|M|_k$ to refer to the sum of the $k$ highest weight edges in $M$, i.e. the value of the top-k objective for $M$.

We almost exclusively consider blowups of graphs in this work. By blowup we specifically mean the following process\todo{Is there a(nother) well-known term for this in the literature?}. First, each vertex is replaced by an empty or complete graph on some arbitrary number of new vertices. Then, for each edge $(u, v)$ of the original graph we add all edges between pairs of vertices $u', v'$, where $u'$ belongs to the empty/complete subgraph that replaced $u$ and $v'$ belongs to the subgraph that replaced $v$. The weights of the new edges can be chosen arbitrarily.

We generally use $\prototype$ to refer to the initial graph and $G$ to refer to the resulting graph of the blowup. We also use the term \emph{prototype} to refer to the initial graph $\prototype$. Moreover, we use the term \emph{blob} to refer to a vertex of a prototype $\prototype$ or to the set of vertices that replaced it during the blowup, interchangeably. Similarly, we use the term \emph{band} to refer to an edge of $\prototype$ or to the set of new edges created from it in the resulting graph. We may also use the term band for the set of edges created while blowing-up a vertex into a clique.

Notationally, to keep things clear, we use $E(B)$ to refer to the set of edges created from a band $B$\todo{minor, but I don't see the difference between $B$ and $E(B)$ -- recall you defined ``band'' already as ``or [...] set of new edges created''}. Moreover, given a set of edges $E$, we use $V(E)$ to refer to the set of vertices incident to least one edge of $E$. We remark that $V(\prototype)$ and $E(\prototype)$ still carry the standard meaning they would for any graph, i.e. they refer to the set of vertices and edges of prototype $\prototype$ viewed as a regular graph before the blowup. Hence, to formally refer to the set of edges in $G$ created by some band in $\prototype$ we will use $E(B)$, where $B \in E(\prototype)$.

In general, when we say that we \emph{guess} some sort of information (for example a set of edges or a vertex-separator), we mean that we try all combinations, possibly under some constraints which will be mentioned or clear from the context. Finally, since we focus on perfect matchings, we will assume that the input graphs (not the prototypes) have $2n$ vertices, i.e. $n$ refers to the size of the PM and not the number of vertices as usual.

%% file: bounded_neighborhood_diversity.tex
We define the neighborhood diversity of a graph below.

\begin{definition}[Neighborhood diversity]\label{def:bounded-neighbourhood-diversity}
We say that a graph $G$ has neighbourhood diversity number $\gamma$ if we can partition its vertices into $\gamma$ sets $V_1, V_2, \dots, V_{\gamma}$ such that for any $u, u' \in V_i$ and any $v \neq u, u'$ we have $(u, v) \in E(G) \iff (u', v) \in E(G)$ and $\gamma$ is the minimum integer for which this partitioning is possible.
\end{definition}

In other words, we can think of the vertices of $G$ as each belonging to one of $\gamma$ \textit{types}, and the existence of an edge only depends on the types of the endpoints. Note here that this is only with regard to the \emph{existence} of edges; their weights can be arbitrarily chosen. For example, $r$-partite complete graphs have $\gamma \le r$. Moreover, a blowup $G'$ of a graph $G = (V, E)$ has $\gamma \le |V|$, regardless of the size of $G'$ and the inclusion or not of the edges ``inside'' the blown-up vertices.

It is important to mention that the neighborhood diversity of a graph can be computed in polynomial time. Additionally, an optimal partitioning of the vertices in different sets according to~\Cref{def:bounded-neighbourhood-diversity} can also be computed in polynomial time~\cite{lampis2012algorithmic}. Thus, we may freely assume from now on that such a partition is given.

\paragraph*{Designing the algorithm via reductions.}
We describe an algorithm (Algorithm \ref{alg:basic}) that runs in $FPT$ time parameterized by both $k$ and $\gamma$ (The running time is polynomial if $\gamma$ is bounded by a constant). The idea is that since all vertices of one type are equivalent in terms of their neighbourhoods, any two matchings that use a fixed number of vertices per type are equivalent as far as extendibility goes. Thus, if one knew the precise number of vertices per type which are incident to the top-k edges in the optimal solution, then they would only need to optimize under this constraint (number per type), since any solution (and thus also the best) would be extendible, as the top-k edges of the optimal solution are extendible.

It turns out that replacing the perfect matching constraint with this ``type'' constraint allows one to solve the optimization problem easily. We describe here how to solve this new problem, which we call \textit{Type-Constrained Maximum Weight Matching}, or \textit{TC-MWM}. The idea is quite simple, we simply add some ``killer'' vertices which we fully connect to each type in order to enforce the type constraints and then run Maximum Weight Perfect Matching.

\paragraph*{The TC-MWM Problem.}
In more detail, let $G = (V_1 \cup V_2 \cup \dots \cup V_{\gamma}, E)$ be a graph of bounded neighbourhood diversity and $(c_1, c_2, \dots, c_{\gamma})$ be a tuple of type constraints. The problem of TC-MWM
 asks us to find a set of edges which are incident to precisely $c_i$ vertices in $V_{i}$ and is of maximum total weight under these constraints. 

 \computationalproblem{Type-Constrained Maximum Weight Matching (TC-MWM)}{An edge-weighted graph $G = (V_1 \cup V_2 \cup \dots \cup V_{\gamma}, E)$ of neighborhood diversity $\gamma$ and a tuple $(c_1, c_2, \dots, c_{\gamma})$ with $c_i \in \left[0, |V_i|\right]$, for all $i$.}{Out of all matchings of $G$ which cover \emph{exactly} $c_i$ vertices of $V_i$ for all $i$, compute one which maximizes the total weight.}
 
 We solve TC-MWM by defining the graph $G'$ where we add vertex sets $K_i$ of cardinality $|V_i| - c_i$, fully connected to $V_i$, with each new edge getting weight 0. We run MWPM on $G'$ and then drop any edges not in $G$ from the result. The construction of $G'$ is exemplified in figure \ref{fig:bounded}.

 
 Note that since the sets $K_i$ must be matched, exactly $|K_i|$ vertices of $V_i$ are reserved for that and the rest, $|V_i| - |K_i| = c_i$, are matched to vertices of $G$. So, any candidate solution to MWPM for $G'$ respects the type constraints. We now need to show that it is also part of a perfect matching in $G$. This follows because we can extend $M^*$ to a perfect matching using the sets $K_i$, as there are precisely $|K_i|$ vertices of $V_i$ unmatched in $M^*$.

\paragraph*{Using TC-MWM to solve TkPM.}
 The final piece of the puzzle is: how does one know the number of vertices per type that the optimal TkPM solution uses, i.e. the vector $(c_1, \dots, c_{\gamma})$? The answer is that we can ``guess'' it, i.e. try all possible options. But how many are there? Note that the top $k$ edges of the optimal TkPM solution will ``touch'' precisely $2k$ vertices. These are distributed in \textit{some} way between the $\gamma$ types. Thus, it suffices to enumerate all solutions of the equation $c_1 + c_2 + \dots + c_{\gamma} = 2k$, which are bounded by a function of $k$ and $\gamma$. Namely, this number equals the number of ways to distribute $2k$ balls into $\gamma$ distinguishable bins, which equals $\binom{2k + \gamma - 1}{\gamma - 1}$. Importantly, this number does not depend on $n$. Note, however, that not all solutions of this equation are relevant; some may produce matchings that are not extendible to a PM in $G$. We simply discard such cases. The pseudocode of the overall algorithm that solves TkPM is given in Algorithm \ref{alg:basic}. A formal analysis of its correctness and runtime can be found in~\Cref{app:bnd}.

\SetCommentSty{textrm}
\newcommand{\inlinecomment}[1]{\hfill \textcolor{green!60!black}{\texttt{\#\ #1}}}
\newcommand{\bigcomment}[1]{\textcolor{green!60!black}{\texttt{\#\ #1}}}

 \begin{algorithm}
 \caption{TkPM for Bounded Neighborhood Diversity}\label{alg:basic}
 \KwInput{$k$, $G = (V_1 \cup V_2 \cup \dots \cup V_{\gamma}, E)$}
 \KwOutput{A PM of $G$ maximizing the top-k objective, $\emptyset$ if none exist}
 $M_{best} \gets \emptyset$

 \For{$(c_1, c_2, \dots, c_{\gamma}) \in [0, k]^{\gamma}$ with $\sum_{i=1}^{\gamma}c_i = 2k$}{

 \For{$i\gets 1$ \KwTo $\gamma$}{
 
    $K_i \gets \{v_{i_1}, v_{i_2}, \dots, v_{i_{|V_i| - c_i}}\}$     
    $E_i' \gets \{ \weightededge{u}{v}{0} \: | \: u \in V_i, v \in K_i\}$ 
    
    }
$G' \gets (V(G) \cup \bigcup_{i = 1}^{\gamma}K_i, E(G) \cup \bigcup_{i = 1}^{\gamma}E_i')$

$M' \gets \textsc{MaximumWeightPerfectMatching}(G') \cap E(G)$

$G'' \gets G[V(G) \setminus V(M')]$

\If{$G''$ has a Perfect Matching $M''$ and $\topkvalue{M'} > \topkvalue{M_{best}}$}{$M_{best} \gets M' \cup M''$}}

\Return $M_{best}$
\end{algorithm}
\subsection{An Approximation Scheme.}
Algorithm \ref{alg:basic} essentially implements a brute force search throughout \emph{all} the options that the optimal solution $M^*$ has for distributing its $2k$ contributing edge endpoints into the blobs of the graph. Hence, for the correct guess of this distribution, we retrieve an optimal solution. What if instead of an optimal solution, we aim for an approximate one? This is in fact possible with slight changes to the algorithm, resulting in an approximation scheme for it. That is, for any desired $\epsilon > 0$, we can output a PM which achieves a $(1 - \epsilon)$ approximation for the top-k objective. We describe~\Cref{alg:approx} below.
\paragraph*{Exponentially growing \emph{band} counts.} 
The first main difference from Algorithm \ref{alg:basic} is that instead of considering the number of vertices touched per blob (the $c_i$ values), we focus on the number of edges utilized per band and also within blobs (we shall use $b_i$ to refer to these numbers)\todo{What would this decomposition yield for the original, non-approximative problem? Would it give anything better or worse than decomposing it by blobs?}. We might slightly abuse terminology (as warned) and use the term band below to also refer to a set of edges staying within a blob\todo{Very confusing, you said earlier that there is no term for edges staying inside a blob, why not create one}. Note that now we may have up to $\binom{\gamma}{2} + \gamma$ different $b_i$ values, whose sum must not exceed $k$. The trick we utilize to achieve the aforementioned approximation scheme is to consider the values $A = \{0, 1, \left \lceil \ASfactor \right \rceil, \left \lceil \ASfactor^2 \right \rceil, \dots, k\}$ for each $b_i$ instead of all values in $[0, k]$, for some appropriate $\ASfactor > 1$ that depends on $\epsilon$. We additionally replace the constraint $\sum_{i=1}^{\gamma}c_i = 2k$ with $\sum_{i=1}^{\gamma}b_i \le k$. Therefore, we might not get the distribution of $M^*$'s edges (per band/within blobs) exactly right, but we get one which is close enough for our purposes (and which is also extendible). More details on this are found in the formal proof.



\paragraph*{Runtime.}
For such a choice of $\ASfactor$, we see that each $c_i$ is endowed with roughly $\log_{\ASfactor}k$ different choices. Hence, we can have at most 
\begin{displaymath}
\bigO\left(\left(\log_{\frac{1}{1 - \epsilon}}k\right)^{\gamma^2}\right) = \bigO\left(\left(\frac{\log_{2}k}{\log_{2}\left(\frac{1}{1 - \epsilon}\right)}\right)^{\gamma^2}\right)   
\end{displaymath}
different combinations for these numbers, which governs the time complexity of our approximation scheme. Notice that we have thus greatly reduced the effect of $k$ on the runtime, introducing an $\epsilon$-dependent overhead instead. In fact, for $\gamma = \bigO\left(\sqrt{\frac{\log{n}}{\log{\log{n}}}}\right)$ we get a \emph{Polynomial-Time Approximation Scheme (PTAS)}, since then for any fixed $\epsilon$ we can bound the overall runtime by a polynomial (whose exponent is roughly the hidden constant in the asymptotic bound for $\gamma$).

A formal analysis of the algorithm can be found in~\Cref{app:approx}.
 \begin{algorithm}
 \caption{Approximation Scheme for TkPM.}\label{alg:approx}
 \KwInput{$k$, $G = (V_1 \cup V_2 \cup \dots \cup V_{\gamma}, E)$, approximation slack $\epsilon \in (0, 1)$.}
 \KwOutput{A PM of $G$ approximating the top-k objective, $\emptyset$ if none exist}
 $M_{best} \gets \emptyset$ 

  $\ASfactor \gets \frac{1}{1 - \epsilon}$\\
  $A \gets \{0, 1, \left \lceil \ASfactor \right \rceil, \left \lceil \ASfactor^2 \right \rceil, \dots, k\}$

 \For{$\left(b_1, b_2, \dots, b_{\binom{\gamma}{2} + \gamma}\right) \in A^{\binom{\gamma}{2} + \gamma}$ with $\sum_{i=1}^{\binom{\gamma}{2} + \gamma}b_i \le k$}{

  Compute the corresponding tuple $(c_1, c_2, \dots, c_{\gamma})$\\

 \For{$i\gets 1$ \KwTo $\gamma$}{
 
    $K_i \gets \{v_{i_1}, v_{i_2}, \dots, v_{i_{|V_i| - c_i}}\}$

    $E_i' \gets \{ \weightededge{u}{v}{0} \: | \: u \in V_i, v \in K_i\}$ 
    }
$G' \gets (V(G) \cup \bigcup_{i = 1}^{\gamma}K_i, E(G) \cup \bigcup_{i = 1}^{\gamma}E_i')$

$M' \gets \textsc{MaximumWeightPerfectMatching}(G') \cap E(G)$

$G'' \gets G[V(G) \setminus V(M')]$

\If{$G''$ has a Perfect Matching $M''$ and $\topkvalue{M'} > \topkvalue{M_{best}}$}{$M_{best} \gets M' \cup M''$}}

\Return $M_{best}$
\end{algorithm}

%% file: recursive_algorithm.tex

\paragraph{Discovering the Algorithm.} Before presenting the new class of graphs we will treat and our algorithm, let us go through the thought process by which one decides upon them. Recall that the graphs of bounded neighborhood diversity that we considered can be simply viewed as arbitrary blowups of ``small'' graphs. Let us use the word \emph{prototype} to refer to these initial graphs. We would like to leverage some sort of structural property (not the property of being small) about these prototypes in order to get an algorithm that runs faster than Algorithm \ref{alg:basic}. The first thought that comes to mind is to consider prototypes of some bounded graph parameter, for example treewidth. Under this assumption, it initially looks like one could repeatedly decompose the problem into smaller subproblems in a Dynamic Programming fashion. However, as will be made clear later, most commonly used parameters such as treewidth or pathwidth fail to give rise to a decomposition scheme for the blown-up graphs. This is due to some very specific demands about the separators of the prototype.\todo{Define either formally or informally what is a separator. What does it mean for a graph with cycles? What if that node is not an articulation vertex and the graph doesn't fall apart?}

\paragraph*{Why blowing-up makes decomposition difficult.} 
Consider some prototype graph $\prototype$. Let us refer to its vertices as \emph{blobs} and to its edges as \emph{bands}. In the resulting graph $G$, the blobs turn into cliques or independent sets and the bands turn into edge sets of a complete bipartite graph. Suppose that we are guaranteed the existence of a small (in terms of number of blobs) separator by some graph parameter. The problem now becomes; how do we actually recurse in order to build an overall solution? It could happen that the small blob-separator of $\prototype$ is actually a huge set of vertices in $G$ which also connects to arbitrarily many edges (for example, see figure \ref{fig:blow_up_difficulty}). Hence, there would be too many different ways to recurse depending on the set of edges chosen to be included in the matching (within the edges separating the graph), defeating the purpose of the decomposition. It seems logical therefore to attempt to avoid this kind of phenomenon and guarantee in some manner that the number of recursion choices is relatively controlled in the blown-up graph.\todo{At this point it is not clear to the reader what decomposition/recursion/etc. you are talking about}


\paragraph*{Avoiding the blow-up side effect.}
How can we mitigate the difficulty explained in the previous paragraph? Our approach is to find a separator which touches a controlled number of (matching) edges in the blown-up graph, hence making the recursion workload less demanding. This motivates the following definitions. Let $M^*$ be the matching of the best solution. The edges of $M^*$ are distributed in \emph{some} way between the bands of the prototype (edges staying within blobs also exist in $M^*$, but that is irrelevant for the following discussion). Some of the bands will contain ``many'' edges, while some will contain ``few''. We characterize the former as \emph{tight} bands and the latter as \emph{loose} bands. The threshold for the number of edges contained in a tight/loose band is set to $\sqrt{n}$. An immediate observation using these definitions is that there can be at most $\sqrt{n}$ tight bands, otherwise there would be more than $n$ edges in $M^*$, a contradiction. An illustration of these definitions is given in figure \ref{fig:loose_tight}. Note that since we are aiming for a subexponential algorithm only, we can assume that we have the information of which bands are tight and which are loose, since we can try every possible combination, of which there are subexponentially many.\todo{Is it subexponential in the total number of nodes in the blown-up graph? You could include a line or two about whether you treat the number of nodes in the prototype constant. The last sentence of the paragraph only seems to hold if the number of nodes in the prototype is sublinear}


\paragraph*{Using the tightness/looseness of bands.}
We now have a tool that we can use to combat the aforementioned blow-up side effect. Specifically, to see why this tool helps, consider the ``lucky'' case where we have a small blob-separator which is incident only to loose bands. Let us call such a separator a \emph{loose separator}. This allows us to make the following optimization in the task of recursing to find a global solution. Instead of considering all subsets of edges within the loose bands, we can simply consider these subsets which respect the bound of $\sqrt{n}$ edges per band, since the optimal solution $M^*$ is guaranteed to fall within such a selection (for the correct choice of tight bands). Of course, we still need to assure that there are few bands also; this is guaranteed by the graph parameter we will use.

\paragraph*{Finding loose separators.}
We have described how in the fortunate case of having a loose separator we can proceed with the decomposition of the problem. The question remains; how do we actually find loose separators? If our graph parameter guarantees the existence of a single small separator (as most commonly used graph parameters do), then the situation looks grim, as it could very well be that some of the blobs are connected to tight bands. A natural idea then is to ask for \emph{many} disjoint small separators. Indeed, if we have this sort of (strong enough) guarantee, we can find one separator which is touched by no tight band (i.e. a loose separator), as each tight band only touches 2 blobs. Therefore, the task is now to find a graph parameter which will guarantee that our prototype graphs have many small separators.

\paragraph*{Graphs of bounded bandwidth have many small separators.}
A parameter such as the one we are looking for turns out to be the bandwidth. The bandwidth of a graph $G$ is defined as follows. 

\begin{definition}\label{def:bandwidth}
    The bandwidth $\bandwidth(G)$ of a graph $G$ is the minimum integer such that there exists an ordering $v_1, v_2, \dots, v_n$ of the vertices of $G$ such that $\{v_i, v_j\} \in E(G) \implies |i - j| \le \bandwidth(G)$.
\end{definition}

It is not very hard to see that any consecutive $\bandwidth(G)$ vertices in such an ordering constitute a separator. Therefore, one can find many disjoint separators of $G$ using that fact. Of course, the separators need to be balanced, in the sense that at least some constant fraction of vertices are preserved in the resulting two parts of the graph after the separation. This can be achieved by e.g. only considering separators of vertices $v_i$ with $i \in [n/4, 3n/4]$. 

Interestingly, the bandwidth serves as an approximate upper bound for the maximum degree. Indeed, there cannot be a vertex $u$ with more than $2\bandwidth(G)$ neighbors, since then in any linear ordering of the vertices there must exist a neighbor $v$ of $u$ which is more than $\bandwidth(G)$ distance apart from $u$ in said ordering, a contradiction. This implies that a blob-separator of small size is also adjacent to a correspondingly small number of bands, which is crucial for our algorithm.

\paragraph*{When to stop separating.}
To find a loose separator, we need to be able to find many small separators compared to the number of tight bands. If the number of separators is more than twice the number of tight bands, there has to exist one which is loose. What if that is no longer true after some steps of decomposition? In such a case, we can show that the number of blobs remaining is of the same order of magnitude as the number of tight bands, i.e.\ we have $\bigO(\bandwidth \sqrt{n})$ blobs. Fortunately, we can now simply use Algorithm \ref{alg:basic} to solve the problem as a base case of the recursion.

\paragraph*{The overall algorithm.}
We described in the previous paragraphs the various ingredients used to make our recursive algorithm. Let us now collect them. First, if the number of tight bands surpasses a certain fraction (depending on $\bandwidth$) of the number of blobs (or if the number of blobs is bounded by some constant), we simply use Algorithm \ref{alg:basic}, since then we would have that the number of blobs is conveniently $\bigO(\bandwidth \sqrt{n})$. Otherwise, the existence of a loose blob-separator of size at most $\bandwidth$ is guaranteed. We find such a separator $\bigseparator$ (it suffices to try all possibilities and does not affect the runtime) and proceed as follows. 

Let $\prototype_{1}$ and $\prototype_2$ be the separated parts of $\prototype$ and $G_1$, $G_2$ the corresponding subgraphs of $G$. We loop through every selection $E'$ of at most $\sqrt{n}$ edges within each band touching $\bigseparator$ such that all edges in $E'$ have at least one endpoint within the vertices in the blobs of $\bigseparator$. For each such $E'$ and for every $(k', k_{blobs}, k_1, k_2) \in [0, k]^4$ such that $k' + k_{blobs} + k_1 + k_2 = k$, we recursively call our algorithm on the inputs $\prototype_1, G_1, k_1$ and $\prototype_2, G_2, k_2$ and join the matchings found with $E'$, as well as a top-$k_{blobs}$ PM calculated for the subgraph of $G$ induced by vertices of blobs of $\bigseparator$ not touched by $E'$ using Algorithm \ref{alg:basic}. If no PM of $G$ can be built in this way, we ignore this specific choice of $E'$. We return the best valid matching found over all such recursive calls. A visual explanation is given in figure \ref{fig:proof_correctness_visual}. The pseudocode is given in Algorithm \ref{alg:recursive}. Its correctness and runtime is proved in~\Cref{app:rec}.


 \begin{algorithm}[H]
 
 \caption{TkPM for Blowups of Bounded Bandwidth Graphs \emph{(BBB)}}\label{alg:recursive}
 \KwInput{$k$, prototype $\prototype$, bandwidth $\bandwidth$, $G$, subset of tight bands $\tightedges \subseteq E(\prototype)$, $n$}
 \KwOutput{A PM of $G$ maximizing the top-k objective, $\emptyset$ if none exist}
\If{$G$ has no Perfect Matching}{\Return $\emptyset$}

$M_{best} \gets \emptyset$

\If{$2|\tightedges| \ge \left \lfloor \left(\frac{|V(\prototype)|}{2} - 3\right) / \bandwidth \right \rfloor
 $ or $|V(\prototype)| \le C$ for some constant $C$}{
    Call Algorithm \ref{alg:basic} on $(k, G)$ and \Return its output.
}
\Else{

Find a separator $\bigseparator$ of $\prototype$ with $V(\tightedges) \cap \bigseparator = \emptyset$ separating $\prototype$ into $\prototype_1$ and $\prototype_2$ with $|V(\prototype_i)| \in \left[\frac{1}{4}|V(\prototype)|, \frac{3}{4}|V(\prototype)|\right]$ for $i \in \{1, 2\}$, such that $|\bigseparator| \le \phi$.

Let $V_{\bigseparator}$ be the set of vertices in blobs of $\bigseparator$.

 \For{$E' \subseteq E(G)$ with $E' \subseteq V_{\bigseparator} \times V(G)$ and $|E' \cap E(B)| \le \sqrt{n} \: \: \: \forall B \in E(\prototype)$}{

  \For{$(k', k_{blobs}, k_1, k_2) \in [0, k]^{4}$ with $k' + k_{blobs} + k_1 + k_2 = k$}{

  $G_i \gets$ the subgraph of $G$ induced by the vertices in blobs of $\prototype_i$ not touched by $E'$, for $i \in \{1, 2\}$\\
  $M_i \gets BBB(k_i, \prototype_i, \bandwidth, G_i, \tightedges \cap E(\prototype_i), n)$, for $i \in \{1, 2\}$

  \If{$M' = E' \cup M_1 \cup M_2$ extends to a PM of $G$}{
  Call Algorithm \ref{alg:basic} on ($k_{blobs}, G_{blobs}$) to get $M_{blobs}$, where $G_{blobs}$ consists only of $V_{\bigseparator}$ and the edges within single blobs of $\bigseparator$.\\
  $M'' \gets M' \cup M_{blobs}$\\
  $M_{best} \gets \max_{|\cdot|_k} \{M'', M_{best}\}$}
  }
 }
}

\Return $M_{best}$
\end{algorithm}

%% file: recursive_EM.tex
Taking a step back, one notices that our recursive approach for TkPM is not very closely related to it but instead exhibits a slight generality. Disregarding the base case, all we did was decompose the problem based on a characterization of the bands using the number of their edges utilized in the solution. Then, we tried all possible combinations of the edges in the loose bands recursively. 

\paragraph*{The EM algorithm.}
In fact, we can use the same approach to derive a subexponential time algorithm for the same class of graphs for EM also. Specifically, we keep roughly the same decomposition procedure and use a different algorithm for the base case. To do this, we rely on the existence of an algorithm solving EM on instances with neighborhood diversity $\gamma$ in time $n^{\bigO(\gamma^{12})}$ from~\cite{maalouly2024exactmatchingproblemdense}. Note however that this algorithm does not run in FPT time, in contrast to our base case algorithm for TkPM. Instead, it runs in XP time, i.e.\ the exponent of the polynomial depends on $\gamma$. We can deal with the consequences of this by tweaking the definition of tight bands to balance the costs of the base case and the recursive decomposition costs.

As we said, we will change the definition of a tight band, specifically by modifying the threshold from $\sqrt{n}$ to $n^{\alpha}$, for some $\alpha < 1$. Note that now this means that the set of tight bands $\tightedges$ can have at most $|\tightedges| \le n^{1 - \alpha}$ elements. This modification affects the runtime of our algorithm in the following two ways:

\begin{itemize}
    \item In the base case, we have that the number of blobs $n'$ is $\bigO(|\tightedges|) = \bigO(n^{1-\alpha})$. So, the runtime using the algorithm mentioned for EM is $n^{\bigO\left(n^{12(1 - \alpha)}\right)}$.
    \item For the recursive decomposition step, we still get the same results about existence of loose separators. However, since loose bands are now allowed to have a different number of edges, we now see that there are at most $\binom{n^2}{n^{\alpha}}^{\bigO(\bandwidth^2)} = n^{\bigO(\bandwidth^2 n^{\alpha})}$ different ways to recurse instead of $n^{\bigO(\bandwidth^2 \sqrt{n})}$.
\end{itemize}

To balance the costs of these two steps, we want to have the same power of $n$ in the exponent, therefore we set $12(1 - \alpha) = \alpha$ and get $\alpha = \frac{12}{13}$. Mimicking the proof of Theorem \ref{the:recursive_runtime}, one sees that with this value of $\alpha$ we get an algorithm that runs in time $2^{\bigO\left(\bandwidth^2 n^{12/13} \log^2{n}\right)}$, which is subexponential.
\recursiveEM*

%% file: appendix.tex
\section{Figures}
\input{figures}
\section{Analysis of Bounded Neighborhood Diversity Algorithm.}\label{app:bnd}
\input{bounded_neighborhood_diversity_proofs}
\section{Analysis of the Approximation Scheme.}\label{app:approx}
\input{approx_scheme_analysis}
\section{Correctness and Runtime of the Recursive Algorithm.}\label{app:rec}
\input{recursive_algorithm_proofs}

%% file: figures.tex
 \begin{figure}
     \centering
     \includegraphics[scale=0.5]{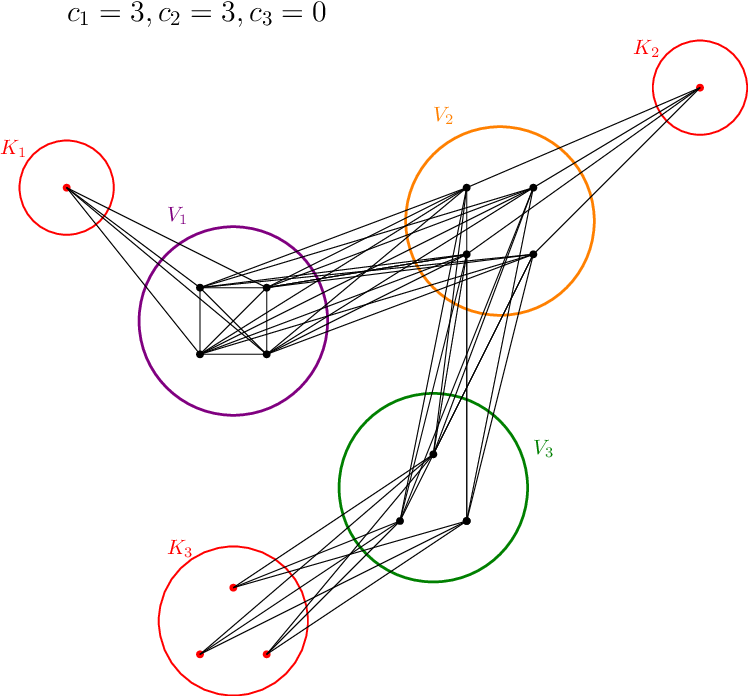}
     \caption{The graph used for the reduction of TC-MWM to MWPM.}
     \label{fig:bounded}
 \end{figure}

 \begin{figure}
     \centering
     \includegraphics[scale=0.5]{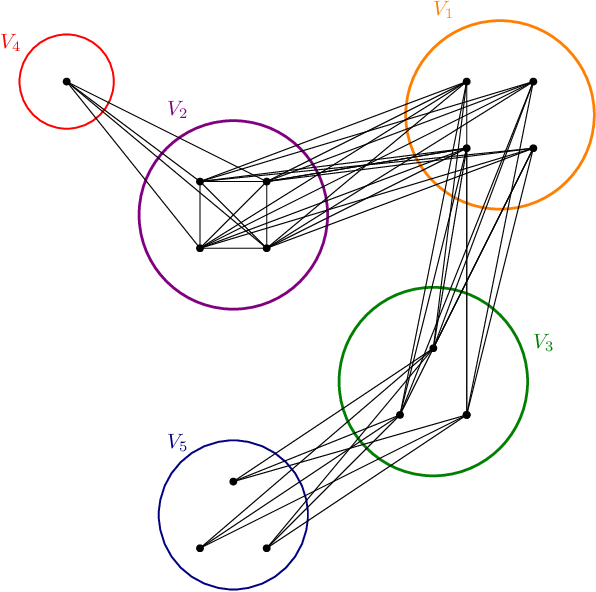}
     \caption{A case where a small separator (here $\{V_1\}$) in the prototype blows up into a big separator in the graph, both in terms of vertices and adjacent edges.}
     \label{fig:blow_up_difficulty}
 \end{figure}

 \begin{figure}
     \centering
     \includegraphics[scale=0.7]{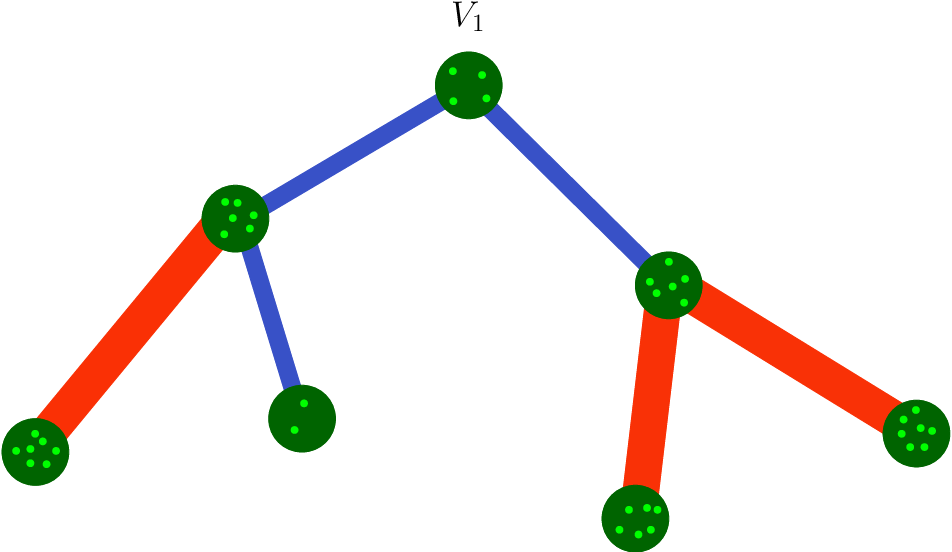}
     \caption{An illustration of loose/tight bands. Loose bands (blue) contain less than $\sqrt{n}$ edges of the solution PM $M$ and tight ones (red) contain at least that many. The set $\{V_1\}$ is a loose separator (see definition \ref{def:loose_separator}), since it splits the prototype in a balanced way, contains at most $\phi$ blobs and is only adjacent to loose bands.}
     \label{fig:loose_tight}
 \end{figure}

 \begin{figure}
     \centering
     \includegraphics[scale=0.7]{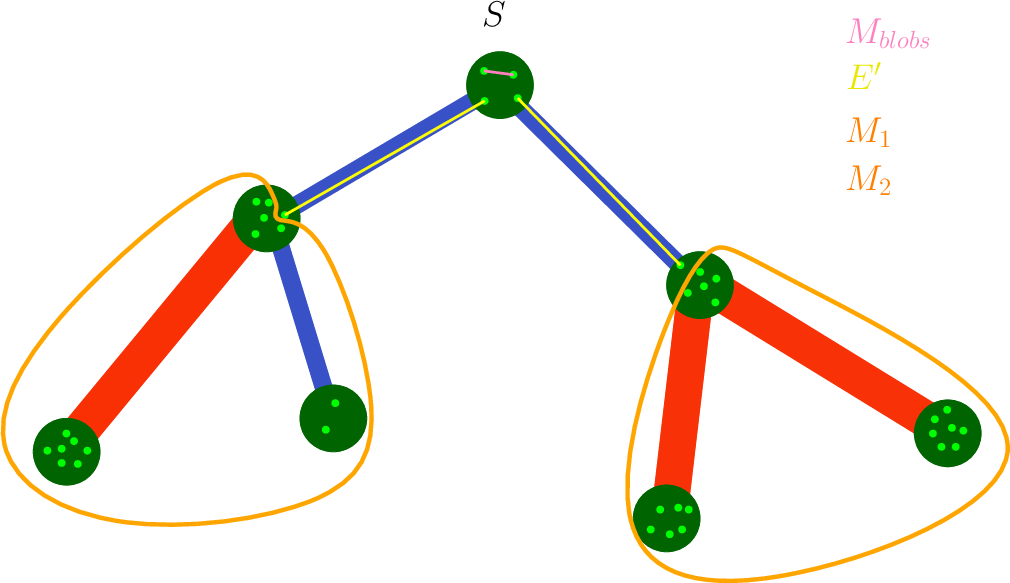}
     \caption{An illustration of the recursive solution to the problem. The complete edge set $E'$ (in yellow) leaving blobs of the loose separator $\bigseparator$ is guessed. Then, the problem decomposes into the matchings $M_1, M_2$ of the smaller subproblems (in orange) as well as the matching $M_{blobs}$ which is comprised of edges which stay within a single blob of $\bigseparator$ (in pink).}
     \label{fig:proof_correctness_visual}
 \end{figure}

 \clearpage

%% file: bounded_neighborhood_diversity_proofs.tex
We now formally prove the correctness and bound the runtime of Algorithm \ref{alg:basic}. We start with a useful lemma regarding the extendibility of partial matchings. The gist of this lemma is the following. If we have two matchings $M_1$ and $M_2$ which are adjacent to (i.e., ``kill'') an equal number of vertices per blob in $G$, then $M_1$ is extendible if and only if $M_2$ is. Note that the weights of the edges are not considered in the lemma below.

\begin{lemma}[Extendibility]\label{lem:extendibility}
    Let $G = (V_1 \cup V_2 \cup \dots \cup V_{\gamma}, E)$ be a graph of neighborhood diversity $\gamma$. Let $M_1$ and $M_2$ be two (partial) matchings of $G$ such that $|V(M_1) \cap V_i| = |V(M_2) \cap V_i|$ for all $i \in [1, \gamma]$. Then, $G[V(G) \setminus V(M_1)]$ contains a Perfect Matching if and only if $G[V(G) \setminus V(M_2)]$ does. In other words, $M_1$ is extendible to a Perfect Matching in $G$ if and only if $M_2$ is.
\end{lemma}

\begin{proof}
    Let $G_i = G[V(G) \setminus V(M_i)]$ for $i \in \{1, 2\}$. We will show that $G_1$ and $G_2$ are isomorphic, which implies the lemma. Indeed, in both $G_1$ and $G_2$, an equal number of vertices remain from each vertex set $V_i$ of $G$, because of the assumption that $|V(M_1) \cap V_i| = |V(M_2) \cap V_i|$ for all $i \in [1, \gamma]$. Let $V_i^1$ and $V_i^2$ be the remaining vertices from $V_i$ in $G_1$ and $G_2$ respectively. Since we have $|V_i^1| = |V_i^2|$, there exists some bijection $f_i$ between these two sets. We fix such a bijection for each $i$ and combine them to get a bijection $f$ from the vertices of $G_1$ to those of $G_2$. Notice that $f$ preserves the type of vertices, hence it is an isomorphism, as the existence of an edge only depends on the types of the corresponding vertices.
\end{proof}
We are now ready to prove the correctness of Algorithm \ref{alg:basic}. We essentially use Lemma \ref{lem:extendibility} to argue that for the correct solution to the equation $\sum_{i=1}^{\gamma}c_i = 2k$, the produced matching (for TCMWM) is indeed extendible to a PM and maximizes the top-k objective.

\basiccorrectness*

\begin{proof}
    If there is no PM of $G$, then we can see that the $M_{best}$ variable will never be updated, hence $\emptyset$ is returned. Otherwise, let $M^*$ refer to the best PM for the top-k objective and let $M_k^* \subseteq M^*$ be the set of edges counting towards the objective. Moreover, let $(c_1^*, c_2^*, \dots, c_{\gamma}^*)$ be the tuple of number of vertices incident to at least one edge of $M_k^*$ in each $V_i$. 

    Consider the iteration of Algorithm \ref{alg:basic} when the tuple $(c_1^*, c_2^*, \dots, c_{\gamma}^*)$ is considered. Then, the variable $M'$ will contain a set of $k$ independent edges of maximum total weight under the constraint that exactly $c_i^*$ vertices from $V_i$ are touched. This implies that $|M'| \ge |M_k^*|$, since $M_k^*$ also respects these constraints. Using Lemma \ref{lem:extendibility}, we conclude that $M'$ is extendible to a PM of $G$, since $M_k^*$ is. Let $M''$ be said extension of $M'$. We now see that when the output is returned, we have
       $$ |M_{best}|_k \ge |M''|_k \ge |M'| \ge |M_k^*| = |M^*|_k,$$
    which concludes the proof.
\end{proof}

We now turn to the runtime of Algorithm \ref{alg:basic}.
\basicruntime*
\begin{proof}
    Indeed, the number of outer loops of the algorithm (each of which incur the cost of constructing $G'$ and running MWPM on it, as well as of checking if $G''$ has a PM) equals the number of solutions to the equation $(c_1, c_2, \dots, c_{\gamma}) \in [0, k]^{\gamma}$ with $\sum_{i=1}^{\gamma}c_i = 2k$ which in turn equals $\binom{2k + \gamma - 1}{\gamma - 1}$, hence the bound follows.
\end{proof}
Although it is not important for our FPT time claim, we remark here that $f(n)$ can be to taken to be $O(|E|\sqrt{|V|})$, hence $O(n^{2.5})$ in the worst case. This is because solving MWPM and checking for the existence of a PM can be done in time $O(|E|\sqrt{|V|})$ by appropriately using the algorithm of Micali and Vazirani~\cite{VaziraniMaximumMatching} for Maximum Matching. For the first task, it suffices to add an appropriately large value to the weights so that the optimal solution is guaranteed to be a PM (for example $n$ times the largest original weight) and for the second one need only put weight 1 on every edge and check that the output is a perfect matching.

%% file: approx_scheme_analysis.tex
\paragraph*{The algorithm and analysis.}
We now proceed to the formal analysis concerning the correctness and runtime of Algorithm \ref{alg:approx}. We first correctness.


\approxcorrecntess*

\begin{proof}
    As usual, let $M^*$ be a PM maximizing the top-k objective. First, we will show that there exists a tuple $\left(b_1', b_2', \dots, b_{\binom{\gamma}{2} + \gamma}'\right) \in A^{\binom{\gamma}{2} + \gamma}$ considered in the algorithm such that a (partial) matching $M'$ is found with $|M'|_k \ge \frac{|M^*|_k}{\ASfactor}$ (where $\ASfactor$ is as defined in the algorithm) and additionally $M'$ uses precisely $b_i'$ vertices from each corresponding band/blob. Next, we will show that $M'$ is extendible in $G$. Since $\ASfactor = \frac{1}{1 - \epsilon}$, it follows that a suitable PM is computed.

    For the first claim, let $\left(b_1^*, b_2^*, \dots, b_{\binom{\gamma}{2} + \gamma}^*\right)$ be the tuple of numbers of edges utilized per band/blob in $M_k^*$, where $M_k^*$ is the set of edges of $M^*$ contributing to the top-k objective. We ignore indices $i$ where $b_i^* = 0$ for the following; these bands/blobs do not provide any weight to the optimal solution (i.e. we focus on iterations where the algorithm gets the zeroes of $b_i^*$ correct). Now, let $b_i'$ be the maximum value in $A$ not exceeding $b_i^*$. We either have that $b_i' = b_i^*$, or that $b_i' = \left \lceil \ASfactor^j \right \rceil$ and $b_i^* < \left \lceil \ASfactor^{j + 1} \right \rceil$, for some $j \ge 0$. In any case we have $\frac{b_i'}{b_i^*} \ge \frac{1}{\ASfactor}$. Now, consider the submatching $M'$ of $M^*_k$ restricted to the best $b_i'$ edges in each band/blob. It follows that $M'$ retains at least $\frac{1}{\ASfactor}$ of the weight of $M^*_k$. Thus, we have $|M'|_k \ge \frac{|M^*|_k}{\ASfactor}$.\todo{Maybe add a sentence about why this argument would not hold (?) if you decompose $k$ like in the original algorithm, only by number of touching edges}

    To see how $M'$ is extendible, we first extend $M'$ to $M''$ by repeatedly adding a suitable edge whenever less than $b_i^*$ edges are used for some band/blob. This is possible since we start with at most $b_i^*$ edges per band/blob. Now, by using Lemma \ref{lem:extendibility} for matchings $M''$ and $M^*_k$, we see that $M''$ is extendible to a PM of $G$, and so the theorem follows.
\end{proof}

Finally, we bound the runtime of Algorithm \ref{alg:approx}.
\approxruntime*
\begin{proof}
    Indeed, there are at most $\bigO(\log_{\ASfactor}k)$ different choices for the value of $b_i$, for each $i$. Hence, at most $\bigO(\left(\log_{\ASfactor}k\right)^{\gamma^2})$ iterations of the outer loop are performed. Since $\ASfactor = \frac{1}{1 - \epsilon}$, the bound follows.
\end{proof}

%% file: recursive_algorithm_proofs.tex
Now we will proceed to formally prove the correctness and runtime bound of Algorithm \ref{alg:recursive}. We start by formally defining loose separators and showing a lemma concerning their existence. Recall that we call a band loose (tight) if less than (at most) $\sqrt{n}$ of its corresponding edges are included in the optimal solution $M^*$.

\begin{definition}[Loose separator]\label{def:loose_separator}
Let $\prototype$ be a prototype of bandwidth $\bandwidth$. Also, let $\tightedges \subseteq E(\prototype)$ be a subset of (tight) bands. We call $\bigseparator \subseteq V(\prototype)$ a \emph{loose separator} if it separates $\prototype$ into $\prototype_1$ and $\prototype_2$ with $|V(\prototype_i)| \in \left[\frac{1}{4}|V(\prototype)|, \frac{3}{4}|V(\prototype)|\right]$ for $i \in \{1, 2\}$, $\bigseparator \cap V(\tightedges) = \emptyset$, i.e. no tight band touches a blob of $\bigseparator$, and additionally $|\bigseparator| \le \bandwidth$.
\end{definition}

An illustration showing a loose separator can be found in figure \ref{fig:loose_tight}.

\begin{lemma}\label{lem:loose_separators}
    Let $\prototype$ be a prototype of bandwidth $\bandwidth$. Also, let $\tightedges \subseteq E(\prototype)$ be a subset of its bands. If $2|\tightedges| < \left \lfloor \left(\frac{|V(\prototype)|}{2} - 3\right) / \bandwidth \right \rfloor
 $, then $\prototype$ contains a loose separator.
\end{lemma}
\begin{proof}
    Let $b_1, b_2, \dots, b_{n'}$ be the linear ordering of $V(\prototype)$ as described in definition \ref{def:bandwidth} (thus $|V(\prototype)| = n'$). Now, consider the following set of candidate separators. We focus on $b_{s}, b_{s + 1}, \dots, b_{e}$, where $s = \left \lceil \frac{n'}{4} \right \rceil + 1$ and $e = \left \lfloor \frac{3n'}{4} \right \rfloor - 1$. Note that the length of this subsequence of vertices is $e - s + 1 = \left \lfloor \frac{3n'}{4} \right \rfloor - \left \lceil \frac{n'}{4} \right \rceil - 1 \ge \frac{n'}{2} - 3$. Moreover, any $\bandwidth$ consecutive vertices in this ordering separate $\prototype$ in the desired balanced manner (the separated parts $\prototype_1$ and $\prototype_2$ consist of the blobs before and after the separator in the entire ordering respectively). There are at least $\left \lfloor \left(\frac{n'}{2} - 3\right) / \bandwidth \right \rfloor$ disjoint such separators. Each band in $\tightedges$ can touch (and thus force us to reject) at most 2 of these separators. Therefore, there must exist one separator $\bigseparator$ among them which is adjacent to no such band (due to the inequality in the lemma statement), which is thus a loose separator.
\end{proof}

\recursivecorrectness*
\begin{proof}
    First of all, when there is no PM of $G$, the theorem is trivially true (the algorithm explicitly checks for it). For the other case, we proceed by induction on the number of blobs, i.e. $|V(\prototype)|$. The base case of the induction (all cases with at most $C$ blobs, where $C$ is some arbitrarily chosen constant) is handled by noting that Algorithm \ref{alg:basic} is called. In a similar way we handle the cases where $2|\tightedges| \ge \left \lfloor \left(\frac{|V(\prototype)|}{2} - 3\right) / \bandwidth \right \rfloor
 $, i.e. using Theorem \ref{the:basic_correctness}. Otherwise, we can see the following.

By Lemma \ref{lem:loose_separators}, there must exist a loose separator in $\prototype$, hence the corresponding search for it in the algorithm succeeds. It now remains to show that for some $E'$ and some combination of $k', k_{blobs}, k_1, k_2$, the variable $M_{best}$ will be updated with a PM $M''$ with $|M''|_k \ge |M^*|_k$, where $M^*$ is the optimal solution. A visual aid to the structure of the rest of the proof can be found in figure \ref{fig:proof_correctness_visual}.

    Indeed, by the assumption that $\tightedges$ is chosen correctly (i.e. $M^*$ contains at most $\sqrt{n}$ edges from each edge set of a band outside $\tightedges$), the restriction of $M^*$ to the edges of bands adjacent to blobs of $\bigseparator$ will constitute one of the choices for $E'$ for the algorithm; we focus on this specific iteration. Moreover, $M^*$ will utilize (with regards to the top-k objective) some number of edges within $E'$ ($k'$), some within single blobs of $\bigseparator$ ($k_{blobs}$) as well as some number of edges within $G_1$ and $G_2$ respectively ($k_1$ and $k_2$). Again, this specific combination of numbers will be considered at some iteration of the inner loop (where we go through the quadruples $k', k_{blobs}, k_1, k_2$) which we now consider for the rest of the proof. Let $M^* = E' \cup M^*_{blobs} \cup M^*_1 \cup M^*_2$, where $M^*_i$ is fully contained within $G_i$ and is a PM for it. Also, $M^*_{blobs}$ is the part of $M^*$ of edges that run within single blobs of $\bigseparator$ (in the case where some blobs were blown up to be complete graphs). By the induction hypothesis, a PM $M_i$ of $G_i$ is returned which maximizes the top-$k_i$ objective, for $i \in \{1, 2\}$. Also, by Theorem \ref{the:basic_correctness}, the variable $M_{blobs}$ contains a PM maximizing the top-$k_{blobs}$ objective in $G_{blobs}$, where $G_{blobs}$ is as in the algorithm pseudocode. Therefore, after this iteration we have
    $$
      |M_i|_{k_i} \ge |M^*_i|_{k_i} \text{ for } i \in \{1, 2\} \text{ and } |M_{blobs}|_{k_{blobs}} \ge |M^*_{blobs}|_{k_{blobs}} \implies$$
      $$|E' \cup M_{blobs} \cup M_1 \cup M_2|_k \ge |E' \cup M^*_{blobs} \cup M^*_1 \cup M^*_2|_k \implies |M_{best}|_k \ge |M^*|_k,
$$
which concludes the proof.
    
\end{proof}

We now turn to the runtime\footnote{It is true that Algorithm \ref{alg:recursive} operates under the assumption that the set of tight bands $\tightedges$ is given. However, as mentioned previously, we can try all combinations for it incurring a multiplicative overhead cost of at most $2^{\bigO \left(\sqrt{n} \log{n}\right)}$, which does not affect the asympotic computational complexity.} of Algorithm \ref{alg:recursive}.

\recursiveruntime*

\begin{proof}
    Let $n' = \bigO(n)$ refer to the number of blobs in $\prototype$. Let $T(n')$ refer to the runtime of Algorithm \ref{alg:recursive}. We can see that for the base case (i.e. when Algorithm \ref{alg:basic} is called), we have $n' = \bigO(\bandwidth\sqrt{n})$ and so the runtime is at most $2^{\bigO \left( \bandwidth \sqrt{n} \log{n} \right)}$ by Theorem \ref{the:basic_runtime}. Otherwise, there exists a function $f(n)$ which is either strictly polynomial (e.g. if $\bandwidth$ is constant) or in any case at most $n^{\bigO(\bandwidth)} = 2^{\bigO(\bandwidth \log{n})}$ (in the case where finding the $\bandwidth$-sized separator $\bigseparator$ is the bottleneck) so that the following holds.

    $$T(n') \le T\left(\frac{3n'}{4}\right) \cdot f(n) \cdot 2^{8\bandwidth^2 \sqrt{n} \log{n}}.$$

    To see why, observe the following. The computed separator $\bigseparator$ has $|\bigseparator| \le \bandwidth$ and moreover the maximum degree of $\prototype$ is at most $2\bandwidth$ (see definition \ref{def:bandwidth}). Hence, there are at most $2\bandwidth^2$ bands to be considered for the recursion step. For each of them, we have to select a subset of at most $\sqrt{n}$ edges. In the worst case, there are $4n^2$ edges to choose from and thus one can bound the number of such edge subset selections by $2^{4\sqrt{n}\log{n}}$ for each band (first pick \emph{exactly} $\sqrt{n}$ edges and then decide which stay). This upper bound is loose for the sake of simplicity but it does not alter the end result asymptotically. To get the final desired bound, we have to raise this number to the power of $2\bandwidth^2$ to account for all combinations of edge subset selections across all considered bands. All other computational costs/recursion choices are hidden within $f(n)$. The $3/4$ fraction for the right hand side stems from the fact that $\bigseparator$ is a loose (and thus appropriately balanced) separator of $\prototype$.

    Having established this inequality, we can now conclude as follows. It takes logarithmically (with respect to $n'$, and therefore also with respect to $n$) many consecutive applications of this inequality so that the argument of $T(\cdot)$ becomes constant (i.e. at most $C$, where $C$ is the arbitrary constant chosen in the algorithm) or the number of tight bands compared to the blobs surpass the threshold for using Algorithm \ref{alg:basic}. Moreover, recall that the base case calls cost at most $2^{\bigO \left( \bandwidth \sqrt{n} \log{n} \right)}$. Therefore, we have\todo{Where do you take into account that you have to guess which band is tight and which band is loose? It adds a multiplicative factor outside, and it can be quite big if there are many small blobs, no?}

    $$T(n') \le 2^{\bigO \left( \bandwidth \sqrt{n} \log{n} \right)} \cdot  \left( f(n) \cdot 2^{8\bandwidth^2 \sqrt{n} \log{n}} \right)^{\bigO(\log{n})} \le 2^{\bigO\left(\bandwidth^2 \sqrt{n} \log^2{n}\right)}.$$
\end{proof}

Interestingly, the runtime remains subexponential whenever $\bandwidth = o\left(\frac{n^{1/4}}{\log{n}}\right)$.